\documentclass[english,prl,a4paper,aps,showpacs,twocolumn,floats]{revtex4}
\usepackage[latin9]{inputenc}
\usepackage{amsmath}
\usepackage{amsthm}
\usepackage{amssymb}
\usepackage{graphicx}

\makeatletter
\@ifundefined{textcolor}{}
{%
 \definecolor{BLACK}{gray}{0}
 \definecolor{WHITE}{gray}{1}
 \definecolor{RED}{rgb}{1,0,0}
 \definecolor{GREEN}{rgb}{0,1,0}
 \definecolor{BLUE}{rgb}{0,0,1}
 \definecolor{CYAN}{cmyk}{1,0,0,0}
 \definecolor{MAGENTA}{cmyk}{0,1,0,0}
 \definecolor{YELLOW}{cmyk}{0,0,1,0}
}
\theoremstyle{plain}

  \theoremstyle{definition}
  
  \theoremstyle{plain}

\usepackage{MnSymbol}
\usepackage{slashbox}
\usepackage{hyperref}

\makeatother

\usepackage{babel}
  \providecommand{\corollaryname}{Corollary}
  \providecommand{\definitionname}{Definition}
\providecommand{\theoremname}{Theorem}

\begin{document}

\title{A Completely Top-Down Hierarchical Structure in Quantum Mechanics}

\author{Yakir Aharonov}

\affiliation{School of Physics and Astronomy, Tel Aviv University, Tel Aviv 6997801,
Israel}

\affiliation{Institute for Quantum Studies and Schmid College of Science and Technology, Chapman University, Orange 92866, CA, US}

\author{Eliahu Cohen}

\affiliation{H. H. Wills Physics Laboratory, University of Bristol, Tyndall Avenue,
Bristol, BS8 1TL, UK}

\email{eliahu.cohen@bristol.ac.uk}

\author{Jeff Tollaksen}

\affiliation{Institute for Quantum Studies and Schmid College of Science and Technology, Chapman University, Orange 92866, CA, US}

\email{tollakse@chapman.edu}

\pacs{03.65.Ta, 03.65.Ca, 03.65.Ud}

\begin{abstract}
Can a large system be fully characterized using its subsystems via inductive reasoning?  Is it possible to completely reduce the behavior of a complex system to the behavior of its simplest ``atoms''? In the following paper we answer these questions on the negative for a specific class of systems and measurements. We begin with simple two-particle example, where strong correlations arise between two  apparently empty boxes. This leads to new surprising effects within atomic and electromagnetic systems. A general construction based on pre- and post-selected ensembles is then suggested, where the $N$-body correlation can be genuinely perceived as a global property, as long as one is limited to preforming a small set of measurements which we term ``strictly local''. We conclude that within time-symmetric quantum mechanics and under certain boundary conditions, high-order correlations can determine low-order ones, but not vice versa. Moreover, the latter seem to provide no information at all regarding the former. This supports a top-down structure in many-body quantum mechanics.
\end{abstract}
\maketitle
An old dispute is well known in philosophical and physical literature between inductive (bottom-up) reasoning versus deductive (top-down) reasoning. The basic question is whether the behaviour of a complex system can be reduced to the behavior of its single elements. After many years of an apparent dominance of reductionism in physics, Anderson defended the opposing view of deductive reasoning, evident in many-body physics, within his seminal paper \cite{Anderson}. Since then, many proponents have further argued for a top-down structure in physics \cite{P1,P2,P3,P4} and across science \cite{P5}, but this topic is still highly controversial \cite{O1,O2}. Within quantum mechanics, the question has also attracted much attention \cite{QH1,QH2,QH3,QH4,QH5}. 

In the following we provide a simple yet fundamental argument in favor of top-down logic. Within the proposed gedankenexperiments it seems that one must adopt a top-down view of quantum mechanics, apparently without a possibility to explain many-particle high-order correlations with the aid of lower-order correlations. We present a family of quantum systems obeying a completely top-down hierarchial structure. For showing that, we shall employ a time-symmetric framework of quantum mechanics known as the two-state-vector formalism (TSVF) \cite{ABL,TSVF,Collapse}. Within the TSVF, quantum systems are described by a forward-evolving (pre-selected) state-vector $|\Psi\rangle$, as well as a backward-evolving state-vector $\langle\Phi|$. The resulting {\it two-state} $\langle\Phi|~|\Psi\rangle$ gives rise to the weak value of any operator $A$
\begin{equation}\label{WV}
\langle A \rangle = \frac{\langle \Phi | A | \Psi \rangle}{\langle \Phi | \Psi \rangle},
\end{equation}
measured weakly \cite{AAV,ACE} in the time interval between pre- and post-selection. When the measured operator is dichotomic, i.e. having only two eigenvalues, and the weak value happens to equal one of the eigenstates, it is assured that had we measured it strongly (projectively), we would have found the same outcome \cite{Dichotomic}. That is, in this special case which will be discussed throughout this paper, the weak and the ``strong'' values coincide. Therefore, although weak values are discussed, the results are quite general and can be put into test by either projective or weak measurements.

One question of interest regarding the two-state is whether it can be inferred by using local tomography based on projective measurements. As shown in \cite{Silva}, this task is possible when discussing pre-selected ensembles, but not when discussing pre- and post-selected ensembles. In the latter case, a larger set of Kraus operators is needed. These results were derived for single particles, but can be easily generalized to multi-partite scenarios. Therefore, broadly speaking, local projective operators are not sensitive to subtle multi-partite correlations within pre- and post-selected ensemble.

In this paper, we will be interested in a smaller set of operators, namely, those that allow ``strictly local'' projective measurements. By that we mean that the measurement is applied not only to a single particle, but also to a single position of this particle. For instance, if the system is composed of two particles, each superposed within two different boxes, then a strictly local measurement would project on a single-particle observable localized within one of the boxes. In light of \cite{Silva}, we know that full tomography of the two-state is impossible using this set of measurements, but in the current work our aim is different. We would like to find out to what extent can high-order multipartite correlations be inferred when employing strictly local measurements and the low-order correlations between them. This analysis gives rise to new phenomena which will be also explored.

The rest of the paper is organised as follows. We first analyze a simple case of strict deductive reasoning within two-particle systems. We later demonstrate this kind of reasoning with the aid of new effects. We then generalize this top-down structure to the $N$-particle case, when correlations seem to emerge only when calculated at the level of the system as a whole.

\section{Two-Particle Systems} \label{Sec:2}

\begin{figure}[tbhp]
 \centering \includegraphics[width=.8\linewidth]{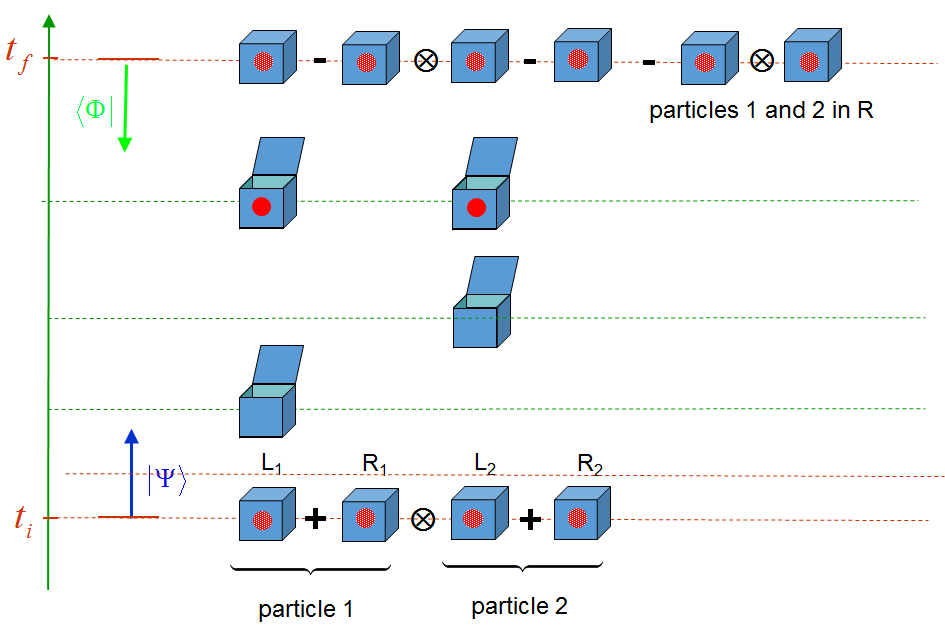}
      \caption{{\bf The basic setup} Two particles are pre- and post-selected. At some intermediate times the particles are separately looked for within the left boxes and then together in a nonlocal manner for finding the correlation between them. The first two measurements imply that the left boxes are empty, while the last measurement reveals a maximal anti-correlation between them.} \label{fig1}
\end{figure}

We begin with a simple illustration of a truly emergent correlation (see Fig. \ref{fig1}). In what follows, we shall describe a two-particle system, where each of the particles can be located in one of two boxes, corresponding to the states $|L_i\rangle$ and $|R_i\rangle$ (right), where $i=1,2$ corresponding to the presence of the $i$ particle in the left/right box. This system is prepared (pre-selected) at time $t=t_i$ in the state:
\begin{equation}
|\Psi\rangle=\frac{1}{2} \left(|L_1\rangle+|R_1\rangle\right) \left(|L_2\rangle+|R_2\rangle \right).
\end{equation}
We later find (post-select) at time $t=t_f$ the system in the state:
\begin{equation}
\langle\Phi|=\frac{1}{\sqrt{3}} \left(\langle L_1|\langle L_2|-\langle L_1|\langle R_2|-\langle R_1|\langle L_2| \right).
\end{equation}


During intermediate times $t_i<t<t_f$, we would like to know whether the particles were in the left boxes. This can be easily checked when calculating the weak values of the projections $\Pi_L^{(i)}\equiv |L_i\rangle\langle L_i|$ for $i=1,2$. According to Eq. \ref{WV} these are:

\begin{equation} \label{Lcorr}
\langle\Pi_L^{(i)}\rangle_w=0.
\end{equation}

That is, if we weakly measure the two left boxes one after the other (e.g. by weakly probing the external fields emerging from these boxes or by a performing a weak scattering experiment), we find that both of them are empty, hence we deduce that both particles were in the right boxes. The same is true for any local property of the particles that we can measure within these two boxes. Moreover, as discussed in the introduction, since projection operators are dichotomic, the inexistence of the particles can be verified with certainty using a ``strong'' (projective) measurement.

Similarly, if we look for the particles in the right boxes, we find them there with certainty:
\begin{equation} \label{Rcorr}
\langle\Pi_R^{(i)}\rangle_w=1.
\end{equation}
The above predictions seem very reasonable, but intriguingly, when calculating the correlation between projections on the left boxes we find that:
\begin{equation} \label{AoH}
\langle\Pi_L^{(1)}\Pi_L^{(2)}\rangle_w=-1,
\end{equation}
and therefore, the two left boxes, which were found to be empty (Eq. \ref{Lcorr}), are in fact correlated in a surprising way.

When we look for the particles in the left boxes, we never find them alone, but we do find a correlation between them. This construction resembles Hardy's thought experiment \cite{Hardy,RHardy,EHardy,Svensson}, but we plan to use it here for a very different purpose. Below we generalize this scenario to the many-body case and claim it is an example for a completely top-down logical structure in quantum mechanics.

One can also calculate the weak values of the other three correlations between the boxes:
\begin{equation} \label{OCorr}
\langle\Pi_L^{(1)}\Pi_R^{(2)}\rangle_w=1,~\langle\Pi_R^{(1)}\Pi_L^{(2)}\rangle_w=1,~\langle\Pi_R^{(1)}\Pi_R^{(2)}\rangle_w=0.
\end{equation}
Such a nonlocal coupling, amenable again to a strong validation, could be verified by joint (nonlocal) measurements or via the recently proposed erasure protocol \cite{RS,BC}.

The nonlocal (multipartite) weak values in Eqs. \ref{AoH},\ref{OCorr} allow to restore in a top-down approach the local (single-particle) weak values of Eqs. \ref{Lcorr},\ref{Rcorr}
\begin{equation} \label{Lcorr2a}
\langle\Pi_L^{(1)}\rangle_w=\langle\Pi_L^{(1)}\Pi_R^{(2)}\rangle_w+\langle\Pi_L^{(1)}\Pi_L^{(2)}\rangle_w=0,~
\end{equation}
\begin{equation}\label{Lcorr2b}
\langle\Pi_L^{(2)}\rangle_w=\langle\Pi_R^{(1)}\Pi_L^{(2)}\rangle_w+\langle\Pi_L^{(1)}\Pi_L^{(2)}\rangle_w=0,
\end{equation}

\begin{equation} \label{Rcorr2a}
\langle\Pi_R^{(1)}\rangle_w=\langle\Pi_R^{(1)}\Pi_L^{(2)}\rangle_w+\langle\Pi_R^{(1)}\Pi_R^{(2)}\rangle_w=1,~
\end{equation}
\begin{equation} \label{Rcorr2b}
\langle\Pi_R^{(2)}\rangle_w=\langle\Pi_L^{(1)}\Pi_R^{(2)}\rangle_w+\langle\Pi_R^{(1)}\Pi_R^{(2)}\rangle_w=1.
\end{equation}
However, all the strictly local information available through the single-particle weak values (Eqs. \ref{Lcorr}, \ref{Rcorr}) is insufficient for calculating in a bottom-up approach the two-particle correlations. This may suggest a true necessity of a top-down logical structure in quantum mechanics,in cases where the experimenters are only allowed to perform strictly local projective measurements. We stress, however, that had we enlarged the set of possible operations, i.e. had we allowed each party to probe its two boxes simultaneously, they could have constructed the two-state and could have revealed the subtle correlations between the boxes \cite{Silva}.

\subsection{A Surprising Manifestation of a Top-Down Structure} \label{Sec:2.1}
Let us imagine a Hydrogen atom located either in box A or box B. We further assume that its electron can occupy the ground state ($|gr\rangle_e$) or the first excited state ($|ex\rangle_e$). We pre-select the atom (separated for convenience into proton, p, and electron, e) as follows:
\begin{equation} \label{Apreselection}
|\Psi\rangle= \frac{1}{\sqrt{3}}\left(|A\rangle_p|gr\rangle_e+|B\rangle_p|gr\rangle_e+|B\rangle_p|ex\rangle_e\right),
\end{equation}
and later post-select on
\begin{equation} \label{Apostselection}
|\Psi\rangle= \frac{1}{\sqrt{3}}\left(|A\rangle_p|gr\rangle_e+|B\rangle_p|gr\rangle_e-|B\rangle_p|ex\rangle_e\right).
\end{equation}
Now, was there a proton in Box B between pre- and post-selection? Apparently not:
\begin{equation} \label{Bproton}
\langle\Pi_B^p\rangle_w=0,
\end{equation}
the weak value of the corresponding projection operator, $\Pi_B^p=|B\rangle_{pp}\langle B|$, is zero (the absence of the proton can be understood as a sum over particle and ``counter-particle'' \cite{CE,ACCE}, i.e. over positive and negative weak values). However, if we look for the pair proton-electron in the ground state within box B, we would find it there with certainty:
\begin{equation} \label{Bpe}
\langle\Pi_B^p\Pi_B^{e,gr}\rangle_w=1.
\end{equation}
Had we trusted only the single-particle weak values, then Eqs. \ref{Bproton},\ref{Bpe} would have suggested that the electron in box B is encircling nothing. To understand why the electron is in a bound state, we must take into account the product of electron and proton observables. Hence we see the fundamental importance of correlations in determining the system's behavior, which goes far beyond the single-particle properties. This is also implied by the possibility to reconstruct the single-particle weak value $\langle\Pi_B^p\rangle_w$ as a sum of two correlations
\begin{equation} \label{Bpepg}
\langle\Pi_B^p\rangle_w=\langle\Pi_B^p\Pi_B^{e,gr}\rangle_w+\langle\Pi_B^p\Pi_B^{e,ex}\rangle_w=1-1=0,
\end{equation}
and from the impossibility to construct correlations from the single-particle weak values (in this case and in general).

\subsection{Novel interference pattern of electromagnetic fields} \label{subSec:2.2}

We conclude this section with one more result stemming from the above construction. Assume now that the two particles are charged. One can measure weakly (from a large distance) the total electromagnetic energy density of the left boxes (supposed now to be spatially separated from the right boxes) without disturbing the particles' states. The energy density will be proportional to:
\begin{equation} \label{EM}
\langle \vec{E}_{tot} ^2\rangle_w=\langle \vec{E}_1 ^2 \rangle_w+\langle \vec{E}_2 ^2 \rangle_w+ 2\langle \vec{E}_1\cdot\vec{E}_2 \rangle_w,
\end{equation}
where $\vec{E}_i$ is the electric field created by the i-th particle. However, since the left boxes are empty, the first two contributions in Eq. \ref{EM} will be zero, and all the energy will emanate from the interference term depending on the correlation between the boxes (Eq. \ref{AoH}).

\section{The $N$-body Scenario} \label{Sec:3}
The above example is, in fact, a special case out of a broad family of pre- and post-selected ensembles all giving rise to a completely top-down logical structure. In these $N$-body systems we will see that the 1-point function, 2-point function,..., $(N-1)$-point function are strictly zero, while the $N$-point function is non-zero. That is, all lower-order correlations cannot reveal the $N$-th order correlations between the constituents of the system. In what follows the details of such construction are presented.\\

Let $N$ spin-1/2 particles be prepared at time $t=t_i$ in the state:
\begin{equation} \label{pre}
|\Psi\rangle=\prod_{i=1}^N |\uparrow_x\rangle_i,
\end{equation}
where $|\uparrow_x\rangle_i=\left(|\uparrow_z\rangle_i+|\downarrow_z\rangle_i\right)/\sqrt{2}$ is the eigenstate of the Pauli-X matrix characterizing particle $i$. If we interpret the operator $\sigma_z$ as a binary position operator \cite{LAC}, then the state given in Eq. \ref{pre} describes $N$ particles, each of which is superposed in two ``boxes'' (the left is characterized by the +1 eigenvalue of $\sigma_z$ and the right by its -1 eigenvalue).

Assume that later, at time $t=t_f$, the particles are post-selected in the state:
\begin{equation} \label{post}
\langle\Phi|=2^{N/2}\prod_{i=1}^N \langle\downarrow_x|_i+C\prod_{i=1}^N \langle\downarrow_z|_i,
\end{equation}
where $C\neq 0$ is some complex number (since we are interested in calculating weak values, the normalization of the pre- and post-selected states does not influence the results).

During intermediate times $t_i<t<t_f$, we shall employ weak measurements to find specific correlations between the particles' positions denoted by projections on the left boxes, e.g. $(1+\sigma_z^{(i)})/2$, being a projection on the left box of the $i$-th particle (later, we shall see in which cases these weak measurements correspond to strong ones).

It turns out that weak measurements of correlations having the form $\prod_{i=1}^N\{[1+\sigma_z^{(i)}]^{b_i}/2\}$, where $b_i\in\{0,1\}$ not all 1, provide {\it null outcomes}, while the outcome of weakly measuring $\prod_{i=1}^N\{[1+\sigma_z^{(i)}]/2\}$ is {\it non-zero}. This means we would never find a particle in any of the left boxes, nor would we find low-order correlations between the particles (when limited to strictly local, projective measurements). Only the $N$-th order correlation is non-vanishing. To summarize the results, we have:
\begin{equation} \label{pred}
\left\langle\prod_{i=1}^N\left[\frac{1+\sigma_z^{(i)}}{2}\right]^{b_i}\right\rangle_w=\begin{cases}
0 & \exists i.~b_i \neq 0 \\
1/C & \forall i.~b_i=1
\end{cases}
\end{equation}
As we noted in the previous sections, due to the ``dichotomic operator theorem'', the absence of low-order correlations can be verified using (counterfactual) ``strong'' measurements. When $C=1$, we can also verify in this way the $N$-th order correlations as the $+1$ eigenvalue of a dichotomic (projection) operator.

Another interesting case is $|C| \ll 1$, where we can observe, using weak measurements only, a very large, robust $N$-th order correlations, even in the absence of all low-order correlations.  The limiting case $C=0$ is forbidden of course because then the pre- and post-selected states become orthogonal.

Using a different terminology, these results also suggest that one cannot perform a full tomography of the two-state using only strictly local projective measurements (and in fact, not even by using correlations between strictly local measurements involving $N-1$ particles or less). However, when viewing the Hilbert space of the $N$ particles as a one-particle Hilbert space of dimension $2N$, then the task is obviously possible in a nonlocal way using the set of $4\cdot(2N)^2$ Kraus operators described in \cite{Silva}.

We note that the suggested top-down structure is very stable in the following sense. First, modifications the relative phase between the terms in Eq. \ref{post}, as well as the value of $|C|$, will not change qualitatively the predictions of Eq. \ref{pred}, suggesting the emergence of high-order correlations. Therefore, even in the presence of noise or unsharp final projective measurement, inserting modifications of the above kind in the post-selected state, we would still expect a similar effect. Furthermore, if for some reason $m$ particles are discarded during the course of the experiment (and thus are not included in the calculation of the various weak values), the top-down logical structure would still persist, and correlations will emerge at the $N-m$ level of the hierarchy.

\subsection{Additional manifestations of emergent correlations}\label{subSec:3.1}
Let $N$ photons be prepared in the state:
\begin{equation}\label{preph}
|\Psi\rangle=\frac{1}{\sqrt{2}}\left(\prod_{i=1}^N |H\rangle_i+|0\rangle\right),
\end{equation}
where $|H\rangle$ denotes the horizonal polarization and $|0\rangle$  is the vacuum state.
Suppose also that these photons are post-selected in the state:
\begin{equation}\label{postph}
\langle\Phi|=\frac{1}{\sqrt{2}}\left(\prod_{i=1}^N \langle V|_i+\langle0|\right).
\end{equation}
We see that the states are not-orthogonal due to the vacuum contribution. Suppose we wish to know the values of circular polarization during intermediate times. Each projection on the clockwise or anti-clockwise circular polarization will rotate the state of the corresponding pre-selected photon towards the post-selection, but only the product of $N$ such rotations will be non-zero. Hence, we see again a situation where all the products of $0\le m<N$ projections are zero, until $m=N$, which is non-zero, suggesting the emergence of the circular polarization in this pre- and post-selected ensemble.

Various thought experiments can be also designed with Fock states too. We could have, for instance, the pre-selected state
\begin{equation}\label{preFock}
\langle\Phi|=\frac{1}{\sqrt{2}}\left(|1,1,...,1\rangle+|0,0,...,0\rangle\right),
\end{equation}
where $|1,1,...,1\rangle$ describes $n$ particles occupying $n$ modes $k_1,...,k_n$. If we post-select on
\begin{equation}\label{postFock}
\langle\Phi|=\langle 0,0,...,0|,
\end{equation}
then the weak values of the annihilation operators $a_{k_l}$ would be null for all $l$, and similarly, any product of at most $n-1$ different annihilation operators would yield null weak values, in contrast to the n-fold product $\prod_{l=1}^{n}a_{k_l}$ which has a non-zero weak value, namely 1. One can imagine many other thought experiments of this kind.

\section{Discussion}\label{Sec:4}
We have seen that in carefully chosen pre- and post-selected many-particle ensembles, strictly local measurements, as well as low-order correlations, provide null results, but when discussing the system as a whole, strong nonlocal correlations emerge. The high-order correlations enable to construct the low-order correlations, but not vice versa. This was shown for the case of two-level systems, but can be trivially generalized. The top-down construction is always possible due to the linearity of weak values, while, as indicated above, the bottom-up approach may fail. These prediction of the TSVF can be verified through projective measurements, and were further shown to be robust under noise and particle loss. They complement a different kind of emergent correlations previously presented in \cite{Pigeon}.

It may be important to emphasize that these cases are very different from those that could be achieved with classical random variables. With a set of $N$ such random variables $\{X_1,...,X_N\}$ it can be easily arranged that the expected values $E[X_1],...,E[X_N]$, and the low-order correlations are all zero, and yet $E[X_1\cdot...\cdot X_N] \neq 0$. However, in the quantum scenarios discussed above, all the outcomes of strictly local measurements are identically zero, rather than the averages, hence locally one cannot have any indication for the existence of particles there, not to mention the possibility of correlation.

The proposed construction thus strengthens a top-down deductive reasoning in many-body quantum mechanics. In addition, it may shed light on some other open problems regarding the foundations of quantum physics, known as the  Oxford Questions \cite{Briggs}, such as ``Does the classical world emerge from the quantum, and if so which concepts
are needed to describe this emergence?'' and ``What experiments are useful for large complex systems, including technological and biological?''.

\section*{Acknowledgements}
We thank Sandu Popescu for very helpful comments and discussions. Y.A. and J.T. acknowledge support
(in part) by the Fetzer Franklin Fund of the John E. Fetzer Memorial Trust. Y.A. also acknowledges support from the Israel Science Foundation (grant no. 1311/14), ICORE Excellence Center ``Circle of Light", and the German-Israeli Project Cooperation (DIP). E.C. was supported by ERC AdG NLST. This research was supported in part by Perimeter Institute for Theoretical Physics. Research at Perimeter Institute is supported by the Government of Canada through the Department of Innovation, Science and Economic Development and by the Province of Ontario through the Ministry of Research and Innovation.


\end{document}